# The Challenges and Opportunities in Creating an Early Warning System for Global Pandemics

**Authors**: David C. Danko[1,4], James Golden[2], Charles Vörösmarty[5,6,7], Anthony Cak[5], Fabio Corsi[5,6], Christopher E. Mason[1,4,10,11], Rafael Maciel-de-Freitas[8,9], Dorottya Nagy-Szakal[3,4], Niamh B. O'Hara[3,4,12]

**Affiliations**:

1. Department of Physiology and Biophysics, Weill Cornell Medicine, New York, NY, USA
2. Rockefeller Foundation, New York, NY, USA
3. Department of Cell Biology, SUNY Downstate Health Sciences University, New York, NY, USA
4. Biotia, New York, NY, USA
5. Environmental Sciences Initiative, Advanced Science Research Center, Graduate Center, City University of New York, New York, NY, USA
6. Department of Geography and Environmental Science, Hunter College, New York, NY
7. Department of Earth and Environmental Sciences, Graduate Center, CUNY, New York, NY, USA
8. Laboratório de Mosquitos Transmissores de Hematozoários, Instituto Oswaldo Cruz, Fiocruz, Rio de Janeiro, RJ, Brazil. CEP: 21040-360.
9. Department of Arbovirology, Bernhard Nocht Institute for Tropical Medicine, Hamburg, Germany. PLZ: 20359
10. The HRH Prince Alwaleed Bin Talal Bin Abdulaziz Alsaud Institute for Computational Biomedicine, Weill Cornell Medicine, New York, NY, USA
11. The Feil Family Brain and Mind Research Institute, Weill Cornell Medicine, New York, NY, USA
12. Jacobs Technion-Cornell Institute, Cornell Tech, New York, NY, USA

**Corresponding author:** niamh.ohara@cornell.edu

The COVID-19 pandemic revealed that global health, social systems, and economies can be surprisingly fragile in an increasingly interconnected and interdependent world. Yet, during the last half of 2022, and quite remarkably, we began dismantling essential infectious disease monitoring programs in several countries. Absent such programs, localized biological risks will transform into global shocks linked directly to our lack of foresight regarding emerging health risks. Additionally, recent studies indicate that more than half of all infectious diseases could be made worse by climate change, complicating pandemic containment. Despite this complexity, the factors leading to pandemics are largely predictable but can only be realized through a well-designed global early warning system. Such a system should integrate data from genomics, climate and environment, social dynamics, and healthcare infrastructure. The glue for such a system is community-driven modeling, a modern logistics of data, and democratization of AI tools. Using the example of dengue fever in Brazil, we can demonstrate how thoughtfully designed technology platforms can build global-scale precision disease detection and response systems that significantly reduce exposure to systemic shocks, accelerate science-informed public health policies, and deliver reliable healthcare and economic opportunities as an intrinsic part of the global sustainable development agenda.

**What is biosurveillance, what is the state of global biosurveillance, and why are we concerned?**

The extraordinary improvement in the physical, social, and economic well-being of humans over the last half-century was stress-tested by a single pathogen: SARS-CoV-2, as it proliferated rapidly across the entire planet. In its wake there followed a profusion of closely coupled societal ills, broken resource supply chains, economic malaise, and an immense failure of public policies whose impacts are felt to this day. No single system can be blamed for the pervasive impacts of SARS-CoV-2, yet early in the pandemic it was clearly difficult for existing public health systems to track the spread of the disease both locally and internationally. As the pandemic progressed, various genomic monitoring systems (such as NextStrain and GISAID) were repurposed to track the evolution and spread of the disease. Though these systems came to life too late to substantially reduce the disease's spread, they were integral to developing effective vaccines, containment strategies, and treatments. Had there been an effective tracking at the start of the pandemic, public health officials would have been able to craft more responsive, informed and mutually-reinforcing policy decisions on travel, lockdowns, mask use, medical supply chains, etc. A pandemic early warning system would therefore have enhanced our collective response capacity and potentially dodged the enormous financial penalty and human death toll associated with the virus.

Despite the clear and present danger, there has yet to be significant coordinated action toward a global biosurveillance and pandemic early warning system (PEWS). In fact, some governments (notably the US and UK) have begun to dismantle the highly successful ad-hoc organizations and data sharing networks developed during the pandemic for testing and epidemiology. Without a dedicated effort to replace these systems there is little reason to think we are collectively more prepared for future pandemics than we were in 2019. And unless such a system is global, it will do little to mitigate the full dimension of the risk.

Indeed, the proliferation of SARS-CoV-2 strains that can evade vaccines, the rise of respiratory syncytial virus (RSV), and an ineffective 2022 influenza vaccine have already given rise to a 'tripledemic' in parts of North America that could have been anticipated far more effectively. A combination of global genetic data to track strains, earth system and human mobility data to model infection rates, and public health statistics to predict disease risk would have greatly improved society's capacity to deal with this compound threat and rebuild public confidence in the effectiveness of healthcare systems that was eroded by COVID-19.

In the longer term, the global economy, environmental stressors, and poor public health management will conspire to intensify pandemic risk. For example, the rise of antimicrobial resistance (AMR) is predicted to cause 10M deaths annually from infection by 2050 (**1**), but little infrastructure exists worldwide to track and predict its spread. Up to 75% of new or emerging infectious diseases are zoonotic in origin (**2**), and these zoonotic "spillovers" frequently occur in biodiverse tropical areas where animal-to-human interactions are co-located with rampant land use change and incursion of human populations (**3**). Climate change will redraw the environmental backdrop within which pandemic risk emerges and elevates the impact of zoonotic infections, which today are responsible for 2.5 billion cases of human illness and 2.7 million human deaths annually (**2**). The lack of basic water services and sanitation affecting billions of people worldwide (**4**) amplifies existing and emerging disease burdens. These interactions demand holistic responses to minimize spillover risk by detecting and then targeting hotspots in which to control deforestation, increase clean water access, regulate animal trading, forecast plausible climate futures, improve the material quality of life, and create targeted genomic tools to inform prediction models of pathogen evolution.

**How a global Pandemic Early Warning system could reduce the risk of pandemics**

Mobilizing digital and social infrastructure will be the key to effective pandemic prevention and response (Figure 1). By connecting a global community of researchers and practitioners, a PEWS integrates signals generated from climate, earth system, biodiversity, and public health surveillance systems. These signals are then used to model risk and provide early warning for emerging infectious disease. If a disease begins to spread, a global PEWS provides the critical platform for initial data sharing to supply unbiased, evidence-driven insights for policy decisions.

Some existing platforms constitute pieces of this infrastructure: GISAID, WHO EWARS, Ecohealth Alliance, and CZID are all data platforms and analytical tools that can track pathogens. However, a global PEWS differs from these previous efforts which were more siloed and single issue-focused. For example, many of these platforms ingest single data streams, such as epidemiological or genomic data to the exclusion of climate and other environmental datasets outlined (Figure 1B). Additionally, these platforms often focus on a single infectious disease threat, instead of monitoring diverse pathogens. Furthermore, with a global problem like pandemics, a global solution is warranted. The challenges to working across nations, including restrictions on data sharing are overcome by a global PEWS using a federated data sharing approach (Figure 1C). In terms of modeling, previous platforms have not leveraged community sourced analysis tools that build the required local relevance (Figure 1D). Finally, previous tools have had limited to no integration with public health agencies (Figure 1F). These many components of a successful PEWS demand an integrated digital infrastructure that draws together groups from multiple countries and sectors, including non-profit, academic, government, and industry, to collect, share, and analyze multi-modal, epidemiological, geospatial, biogeographical, and biodiversity datasets.

**PEWS as a global community health model**

We outline the components of a cross-sector, multi-modal global PEWS framed around *data philanthropy* and *modern data logistics*. Our vision relies on creation of a global network of researchers---a digital collaboratory—connected through an open-source data science platform, representing a globally unique resource. While talent is global, climate and health resources are not, and thus resource-limited sites could greatly benefit from a fully extensible network of tools and methods.

Dramatic improvements in computational technology, access to growing numbers of digital data resources, and increasingly sophisticated Artificial Intelligence (AI) and Machine Learning (ML) algorithms have catalyzed significant breakthroughs in quantitative research over the last 30 years (e.g., **5**). For biosurveillance, AI and ML are of similar importance and relevance. The potential power of AI to support global public health is driven by three important trends:

1) The emergence of large, complex data sets spanning hundreds of thousands to billions of human interactions within society and with climate and environment;

2) Affordable access to high performance computing capabilities, such as the Cloud and advanced graphical processing chips, and new statistical techniques such as Deep Learning; and,
3) The growth of global talent pools with the skills to apply novel techniques to large datasets, the ability to translate results into concrete actions, and the ability to carry out those concrete actions.

Effective pandemic prevention requires prediction of regularities emerging from complex, self-organizing, nonlinear systems. These regularities are found in real-world data, across many scientific and commercial domains, creating both strong and weak signals—there for us to find if we know where, when, and how to look. Using AI, teams of researchers can discover these signals and assemble them into robust models. These models in turn create predictions that can be used to reduce uncertainty and enable creative interventions.

To develop an effective platform for disease prediction we propose the idea of a *community model,* patterned after successful examples drawn from the climate change research community (**6**), quantitative investment firms (**7**), and large-scale cancer research challenges (e.g., DREAM challenges). The core distinction from current non-community driven approaches is that many globally distributed teams can each develop data science applications for specific use cases, compared to a centralized team developing more general but closed models. For example, rather than building a single large model that predicts global weather-related conditions linked to disease transmission from satellite data, the PEWS approach uses distributed teams to create smaller models that will predict endpoints like the temperature in Brazil, chance of rainfall in New York, humidity in Hong Kong, etc. These smaller models are inherently more transparent, can be developed more quickly from targeted data sources, and can be more easily integrated to local response. To predict more complex and interacting phenomena, modeling teams can layer small models together to form larger constellations of linked algorithms, which can be automated using ML techniques.

The foundation of this approach is globally distributed teams of top quantitative researchers, who continuously leverage a growing, pan-signal library to build ensemble models that are resilient to situational change and evolve against new data. The resulting models reflect wider, more granular view of the world than any single descriptive model, and ensemble models already have been successfully used for improved pathogen detection (**8**). In an ideal scenario, researchers would share data and models broadly in open libraries that can be accessed globally. We accept that data cleansing and curation is a key factor that must be addressed. The data and model libraries thus could be connected to other groups and institutions via interoperable federated APIs and researchers could then use diverse datasets without struggling to learn new interfaces and by benefiting from improved crowd-sourced modeling.

**How can the impact of a global PEWS be maximized?**

Prediction lies at the heart of decision-making under uncertainty. Our vision for a global PEWS is as an open-sourced, globally distributed data science platform that enables a global quantitative approach to modeling infectious disease that includes:

- <u>A framework for global data philanthropy</u>. To create effective pandemic mitigations, researchers and policymakers will need access to large, diverse data sets from both public and private sources;

- <u>A computational environment for "idea arbitrage"</u>. A diverse group of globally distributed data scientists will bring non-correlated perspectives to infectious disease data science and remove inherent bias from the system;
- <u>Tools based on Federated learning and data sharing</u>. To preserve data sovereignty and healthcare privacy across participating nations and institutions;
- <u>Seamless extraction of candidate signals</u>. From multiple diverse and differentiated data sources;
- <u>Computational approaches for signal detection</u>. To explore the landscape of all possible prediction models through advanced machine learning; and,
- <u>Forwarding the notion of model resiliency</u>. With a mix of data, tools, methods, and people.

Because pandemics cross borders and affect all sectors of the global ecosystem, the essential need is for cross-national and cross-sectoral collaboration. The COVID-19 pandemic laid the foundation for unprecedented global scientific collaboration, which can now be further supported and accelerated. Expansion of these efforts by establishing an open-sourced, digital platform for a number of emerging infectious diseases and antimicrobial resistant infections is necessary to combat emerging pathogens. In this sense, a multi-faceted global PEWS provides the quintessential example of a global partnership (SDG-17) as envisioned by the Goals' U.N. framers and thus becomes an important lynchpin in the larger sustainable development agenda and security in the modern world. Given the imminent danger of yet another pandemic with unforeseen global consequences, establishing a global pandemic early warning system now takes on renewed urgency. Without a tangible means for prediction, early detection, and continuous monitoring of the next global pandemic, we risk stumbling again as we did through the recent past.

**Figure 1.** Overview of a global platform for monitoring and predicting infectious disease outbreaks (Global PEWS), with an initial use-case for dengue virus in Brazil. This platform quantifies the drivers of disease outbreak (Panel A) by simultaneously ingesting multi-omic datasets, climate, land use, epidemiological, and microbial genomics data (B) to create the foundation for predictive modeling. This new platform differs from previous efforts to monitor pathogens. which only focused on independent data streams, like epidemiological or genomic data, and have supported limited components of the overall system, such as the federated sharing of data only (C), some fragmentary international collaboration (D), and first-order risk maps (E). Furthermore, previous platforms have neither leveraged community sourced analysis tools to build local relevance, nor (F) translated outcomes of PEWs to a general audience and public health officials. The PEWS explicitly integrates these previously distinct components while extending their individual capabilities (A-F).

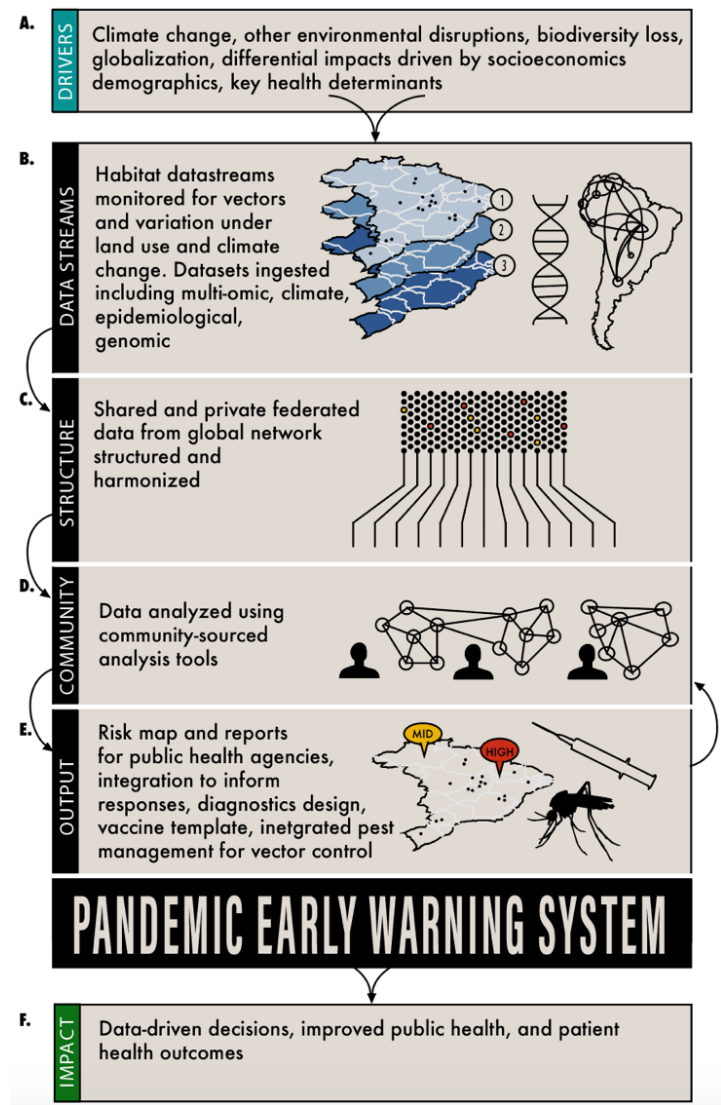